\documentclass[aps,prd,reprint,preprintnumbers,nofootinbib]{revtex4-1}
\pdfoutput=1

\usepackage{color}
\usepackage{graphicx}
\usepackage{amsmath}
\usepackage{amssymb}

\usepackage[colorlinks]{hyperref}
\hypersetup{allcolors=[rgb]{0,0,1}}

\newcommand{\orcid}[1]{\href{https://orcid.org/#1}{#1}}

\usepackage{scalerel}
\newlength{\heightnu}
\AtBeginDocument{\settoheight{\heightnu}{$\nu$}}

\usepackage[normalem]{ulem}

\newcommand{\mn}{\Delta m_{21}^2}
\newcommand{\mt}{\Delta m_{31}^2}

\def\lsim{\raise0.3ex\hbox{$\;<$\kern-0.75em\raise-1.1ex
\hbox{$\sim\;$}}}
\def\gsim{\raise0.3ex\hbox{$\;>$\kern-0.75em\raise-1.1ex
\hbox{$\sim\;$}}}

\keywords{Neutrino Physics, Reactor Experiments}

\begin{document}

\vspace*{-3cm}
\title{  Constraints on the Solar $\Delta m^2$ using Daya Bay \& RENO }

\author{Seon-Hee~Seo}
\email{sunny.seo@ibs.re.kr}
\thanks{\orcid{orcid \# 0000-0002-1496-624X}}
\affiliation{Center for Underground Physics, Institute for Basic Science, 55 Expo-ro Yuseong-gu, Daejeon 34126, Korea \\ }

\author{Stephen~J.~Parke}
\email{parke@fnal.gov} 
\thanks{\orcid{orcid \# 0000-0003-2028-6782}}
\affiliation{Theoretical Physics Department, Fermi National Accelerator Laboratory, P.\ O.\ Box 500, Batavia, IL 60510, USA \\ }
\preprint{FERMILAB-PUB-18-392-T}

\vglue 1.6cm

\vspace*{2.cm}

\begin{abstract}

We demonstrate that the currently running short baseline reactor experiments, especially Daya Bay, can put a significant upper bound on
 $\Delta m^2_{21}$. This novel approach to determining $\Delta m^2_{21}$ can be performed with the current data of both Daya Bay \& RENO and provides additional 
information on $\Delta m^2_{21}$ in a different $L/E$ range ($\sim$ 0.5 km/MeV) for an important consistency check on the 3 flavor massive neutrino paradigm.
Upper limits by Daya Bay and RENO and a possible lower limit from Daya Bay, before the end of 2020,
will be the only new information on this important quantity until the medium baseline reactor experiment, JUNO, 
gives a very precise measurement  in the middle of the next decade. 
In this study $\theta_{12}$ value is fixed since its impact on the $\Delta m^2_{21}$ measurement is relatively small as discussed in the Appendix.

\end{abstract}

\date{September 17, 2018}

\pacs{14.60.Lm, 14.60.Pq }

\maketitle

\section{Introduction}

The fact that neutrinos have mass and mix is now well established by a large number of experiments. In this paper we concentrated on the mass difference squared between the two mass eigenstates that have the most electron neutrino, $\nu_1$ and $\nu_2$.  The splitting between these two neutrinos, $\Delta m^2_{21}\equiv m^2_2-m^2_1$, is responsible for the (anti-) neutrino oscillations observed at an L/E = 15~km/MeV and for the neutrino flavor transformations inside the Sun, hence the name the solar mass squared difference.

In this paper, we demonstate that the currently running 
short baseline  ($\sim$1.5~km)  reactor anti-neutrino experiments, Daya Bay  \cite{An:2015rpe} and RENO \cite{RENO:2015ksa} both  have enough data already collected ($>$ 2,000 days) to constrain  $\Delta m^2_{21}$ to be less  than 3 times the KamLAND central value ($7.5 \times 10^{-5}$ eV$^2$).  By the end of the running time of these experiments, sometime in 2020, they will  be able to constrain this parameter to less than  twice the KamLAND value.  Setting a lower limit maybe possible for the Daya Bay experiment with improvements on their systematic uncertainties. 
Upper, and maybe lower, limits from Daya Bay and RENO, 
will add independent information to our knowledge of  $\Delta m^2_{21}$  and provide an important consistency check of the 3 flavor massive neutrino paradigm.
While not capable of directly addressing the  $\sim$2$\sigma$ tension between KamLAND \cite{Gando:2010aa} reactor experiment ($L/E \sim$ 50 km/MeV)  and the combined Super KamiokANDE \cite{Abe:2010hy} \& Sudbury Neutrino Observatory  \cite{Aharmim:2011vm} solar neutrino measurements of $\mn$,  measurements of $\mn$ by Daya Bay and RENO are at a different $L/E$ range ($\sim$ 0.5 km/MeV) than previous measurements. Furthermore, the ratio of  $\Delta m^2_{21}$  to   $\Delta m^2_{31}$, at  $L/E \sim$ 0.5 km/MeV,  is needed by the long baseline $\nu_e$ appearance experiments for the precision measurement of leptonic CP violation.

Currently the best measurement of the solar mass squared difference, $\Delta m^2_{21}$, is from the long baseline reactor anti-neutrino experiment, KamLAND, which has determined
\begin{eqnarray}
\Delta m^2_{21}  = 7.50^{\,+0.20}_{\,-0.20} \times 10^{-5} ~{\rm eV}^2,
\label{eq:kamland}
\end{eqnarray}
see \cite{Gando:2010aa}.
The only other measurement of $\Delta m^2_{21}$ comes from a combined measurement using the solar neutrino experiments principle Super KamiokaNDE (SK) and Sudbury Neutrino Observatory (SNO). This combined measurement is 
\begin{eqnarray}
\Delta m^2_{21}  = 5.1 ^{\,+1.3}_{\,-1.0}  \times 10^{-5} ~{\rm eV}^2,
\label{eq:sk+sno}
\end{eqnarray}
from SNO \cite{Aharmim:2011vm}. Similar results can be found in SK   \cite{Abe:2010hy} and Nu-Fit \cite{Esteban:2016qun}.
This solar neutrino determination of $\Delta m^2_{21}$ comes from the non-observation of the low energy up turn of the $^8$B neutrino survival probability by both SNO and SK and the observation of a day-night asymmetry by SK. 

CPT invariance implies that the $\mn$ measured in reactor anti-neutrinos and solar neutrinos should be identical. However, at the 2$\sigma$ level there is some tension between these two determinations of this important quantity.   This tension could arise from a statistical fluctuation, some error in the analysis of one or more of the experiments or new physics.

Moreover, $\Delta m^ 2_{21}$ is an important parameter for the determination of the CP-violating phase, $\delta$, in the  long baseline neutrino\footnote{In the rest of this paper, when referring to neutrinos, we mean neutrinos and/or anti-neutrinos.} oscillation experiments (T2K \cite{Abe:2011ks}, NOvA \cite{Ayres:2004js}, DUNE \cite{Acciarri:2015uup}, T2HK \cite{Abe:2015zbg}, T2HKK \cite{Abe:2016ero}) as the size  of the CP violation is proportional to $\Delta m^2_{21}$, as well as other parameters.  In vacuum, at the first oscillation peak, $L/E \sim$ 0.5 km/MeV, for $\nu_\mu \rightarrow \nu_e$:
\begin{eqnarray}
 P(\bar{\nu}_\mu \rightarrow \bar{\nu}_e)-P(\nu_\mu \rightarrow \nu_e) \, \approx \, \pi \, J \, \left( \frac{\Delta m^2_{21}}{\Delta m^2_{31}}\right)
\end{eqnarray}
where $J  =  \sin 2 \theta_{12} \sin 2 \theta_{13} \cos \theta_{13} \sin 2 \theta_{23} \sin \delta \, \approx 0.3  \sin \delta $ is the Jarlskog invariant.

T2K's data point  in the bi-event plane, see Fig. 44 of \cite{Abe:2017vif}, 
\begin{eqnarray}
N(\nu_\mu \rightarrow \nu_e)=37 \quad {\rm and} \quad N(\bar{\nu}_\mu \rightarrow \bar{\nu}_e)  = 4  \nonumber
\end{eqnarray}
being outside the allowed region (by about 1 $\sigma$) could be caused by $\Delta m^2_{21}$ being larger than KamLAND value, twice the KamLAND central value works well. Again, it is probably a statistical fluctuation but with only one precision measurement of   $\Delta m^2_{21}$, other possibilities are not completely excluded.

The future medium baseline reactor experiment JUNO ($L/E \sim$ 15 km/MeV) will measure $\Delta m^2_{21}$ and $\sin^2 \theta_{12}$ with better than 1\% precision, \cite{An:2015jdp}. However, this experiment is under construction and the precision measurements of the solar neutrino oscillation parameters will not be available until approximately 5 years from now.  In more than a decade from now, the DUNE \& HyperK proposed experiments will make a precise measurement of $\mn$ using solar neutrinos, see \cite{Beacom:2018xyz} and \cite{Abe:2018uyc} respectively.

In section II, we discuss in detail the effects of changing $\mn$ on the oscillation probability.  Then in section III we explain and give the results of a simulation of both Daya Bay and RENO using 3000 live days of data with and without systematic uncertainties followed by a conclusion.

\section{Oscillation Probability}

The electron antineutrino  disappearance probability, in vacuum, can be written as
\begin{eqnarray}
& & P(\bar{\nu}_e \rightarrow \bar{\nu}_e )  =  1-P_{13}-P_{12} \quad {\rm with}  \\[2mm]
P_{13} & = & \sin^2 2 \theta_{13}\, ( \cos^2 \theta_{12} \sin^2 \Delta_{31}+\sin^2 \theta_{12} \sin^2 \Delta_{32}),  \nonumber  \\[2mm]
P_{12} & = & \sin^2 2 \theta_{12} \cos^4\theta_{13}  \sin^2 \Delta_{21},  \nonumber
\end{eqnarray}
where $\theta_{12}\approx 33^\circ $ and  $\theta_{13} \approx 8^\circ $ are the solar and reactor mixing angles respectively and 
the kinematic phases are given by $\Delta_{jk} \equiv \Delta m^2_{jk} L/(4E)$.  The $P_{13}$ term  is associated with the atmospheric oscillation scale of 0.5 km/MeV, and the $P_{12}$ term is associated with the solar oscillation scale of 15 km/MeV. 

Using typical fit values and considering a $L/E$ range  around the first oscillation minimum  ($L/E = 0.5\,{\rm km/MeV} $), we can approximate $P_{13}$ and $P_{12}$ as follows:
\begin{eqnarray}
P_{13} & \approx & 0.08 \sin^2 \left( \frac{\pi}{2} \left(\frac{L/E}{0.5 \,{\rm km/MeV}}\right) \right)  \\[2mm]
P_{12} & \approx & 0.002  \left(\frac{L/E}{0.5 \,{\rm km/MeV}}\right)^2  \left(\frac{\Delta m^2_{21}}{7.5 \times 10^{-5} \,{\rm eV^2}}\right)^2.
\end{eqnarray}
 
For  $ \Delta m^2_{21}= 7.5 \times 10^{-5} \,{\rm eV^2}$, the $P_{12}$ term is essentially negligible for all  $L/E < 1\,{\rm km/MeV}$. 
This encompasses the $L/E$ range of all current short baseline experiments.

  However, consider the case that  $ \Delta m^2_{21}$ is 3 times larger than this value, i.e. $ 22.5 \times 10^{-5} \,{\rm eV^2}$, then 
\begin{eqnarray}
P_{12} & \approx & 0.02  \left(\frac{L/E}{0.5 \,{\rm km/MeV}}\right)^2  \left(\frac{\Delta m^2_{21}}{22.5 \times 10^{-5} \,{\rm eV^2}}\right)^2 .
\end{eqnarray}
$P_{12}$ is now no longer negligible compared to  $P_{13}$ at oscillation minimum ($L/E =0.5\,{\rm km/MeV}$) and $P_{12}$ gets larger for  $L/E > 0.5\,{\rm km/MeV}$ whereas $P_{13}$ is getting smaller. In fact, at   $L/E = 1\,{\rm km/MeV}$, $P_{12}$ would be as large as $\sin^2 2 \theta_{13}$ (0.08) for this value of  $ \Delta m^2_{21}$.

Therefore the short baseline reactor experiments can constrain $ \Delta m^2_{21}$ to be less than 2 to 3 times the current best fit value depending on the experiment, Daya Bay or RENO,  run time and the confidence level. Setting a lower bound on  $ \Delta m^2_{21}$ will be challenging for these experiments due to systematic uncertainties.
As data above $L/E  \sim 0.5\,{\rm km/MeV}$ is important for this constrain, the  Double Chooz experiment, which has no data with $L/E > 0.5\,{\rm km/MeV}$, is not considered.

Since the position of the first oscillation minimum for $P(\bar{\nu}_e \rightarrow \bar{\nu}_e)$ is given by
\begin{eqnarray}
\frac{L}{E} \approx \frac{2\pi}{\Delta m^2_{ee}},
\label{eq:mina}
\end{eqnarray}
where $\Delta m^2_{ee} \equiv  \cos^2 \theta_{12} \Delta m^2_{31} +  \sin^2 \theta_{12} \Delta m^2_{32}$ (at least for small $\Delta m^2_{21}$), it is natural to write the disappearance probability in terms of  $\Delta m^2_{ee}$ and  $\Delta m^2_{21}$ as follows, \cite{Nunokawa:2005nx} \& \cite{Parke:2016joa}:
\begin{eqnarray}
& & 1- P(\bar{\nu}_e \rightarrow \bar{\nu}_e)  \approx   
 \cos^4\theta_{13} \sin^2 2 \theta_{12} \sin^2 \Delta_{21} \nonumber  \\ & & 
 + \sin^2 2 \theta_{13} \, \biggr[\, \sin^2 |\Delta_{ee}| + 
    \sin^2 \theta_{12} \cos^2 \theta_{12} \Delta^2_{21}  \cos(2\Delta_{ee}) \nonumber \\
& &   -~\frac{1}{6} \cos 2\theta_{12} \sin^2 2\theta_{12}~\Delta^3_{21} \sin(2\Delta_{ee}) 
+ ~{\cal O}(\Delta^4_{21})~ \biggr]. \label{eqn:taylor}
\end{eqnarray}
For $\Delta_{21} < 0.5$, only the first two of the terms of RHS of eq. (\ref{eqn:taylor}) are larger than 0.005 and  therefore relevant for the analysis\footnote{For small $\Delta_{21}$, the disappearance probability depends on only three variables; $\sin^2 \theta_{13}$, $\Delta m^2_{ee}$ and the combination $\mn \, \sin2\theta_{12}$, see Appendix A.}. Since the experiments of interest, Daya Bay and RENO, have an L/E $< ~1$ km/MeV, the $\Delta_{21} < 0.5$ constraint corresponds to a
$\Delta m^2_{21} <  4 \times 10^{-4}$ eV$^2$ or  5 times the KamLAND value of 7.5 $\times 10^{-5}$ eV$^2$.  Using additional terms of eq. (\ref{eqn:taylor}) will extent the range of applicability.

For small values of $L/E$ ($<$ 0.2 km/MeV), where there is large statistics from the near detectors,
 \begin{eqnarray}
1-P(\bar{\nu}_e \rightarrow \bar{\nu}_e) & &  \nonumber \\
& &  \hspace*{-1.5cm}  \approx   \left[\sin^2 2\theta_{13} +\sin^2 2\theta_{12}  \cos^4 \theta_{13} 
 (\Delta m^2_{21}/\Delta m^2_{ee})^2  \right] 
 \nonumber \\ & & \times 
 (\Delta m^2_{ee} L/4E)^2.
\end{eqnarray} 
To keep the disappearance probability the same as we vary $\Delta m^2_{21}$, at these small $L/E$, we must keep the quantity in $[\cdots]$ in the above equation unchanged.
If we also keep the position of the first minima fixed by holding $\Delta m^2_{ee}$ fixed (see eq. (\ref{eq:mina})), then 
\begin{eqnarray}
& & \sin^2 \theta_{13}+ \sin^2 \theta_{12} \cos^2 \theta_{12}(\Delta m^2_{21}/\Delta m^2_{ee})^2 \nonumber \\
& & \hspace*{2cm} = {\rm constant} \approx 0.021 \nonumber  \\[2mm]
{\rm or} ~~& &  \sin^2 \theta_{13} \approx 0.021 -  2 \times 10^{-4} \left( \frac{\Delta m^2_{21}}{7.5 \times 10^{-5}\, {\rm eV}^2} \right) ^2~~   \label{eq:fix2}
\end{eqnarray}
to leading order in $\sin^2 \theta_{13}$.  So as we vary $\Delta m^2_{21}$ from KamLAND value of $7.5 \times 10^{-5}$eV$^2$, we must also change $\sin^2 \theta_{13}$ from $0.021$ so as to keep the combination in eq. (\ref{eq:fix2}) unchanged.

\begin{figure}[t]
\begin{center}
\includegraphics[width=0.45\textwidth]{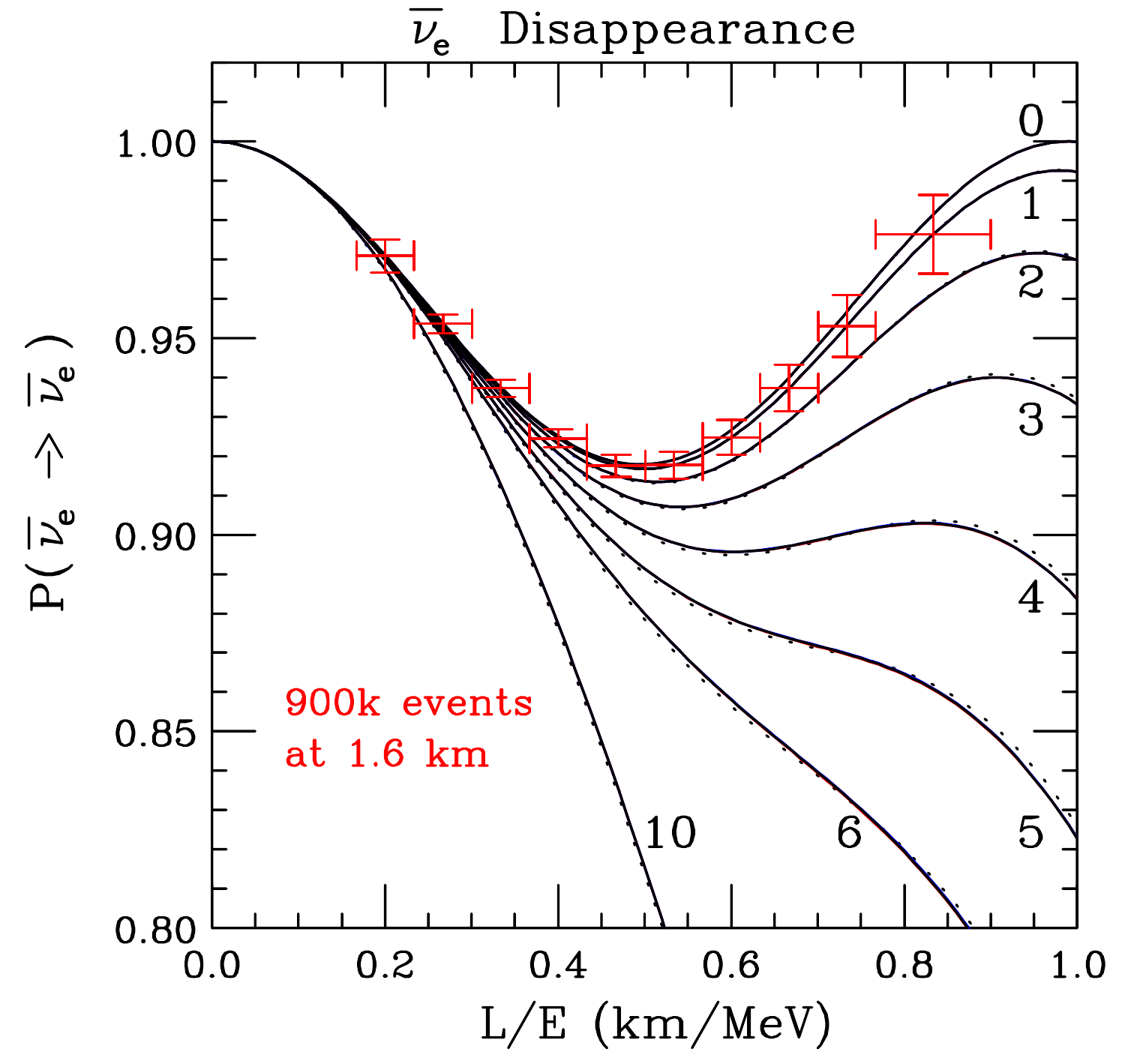}
\end{center}
\caption{The exact electron anti-neutrino disappearance probability (solid) as a function of $L/E$ as $\Delta m^2_{21}$ is varied in multiples (lines labelled = (0, .., 6, 10)) of the KamLAND value of $7.5 \times 10^{-5}~$ eV$^2$. $\theta_{13}$ is also varied,  see eq. (\ref{eq:fix2}), to keep the same disappearance probability for $L/E < 0.2$ km/MeV.  The red points with error bars, are the statistical uncertainties  only, for a detector at 1.6~km from a reactor core with an exposure such that there are 900k events in this detector  assuming the KamLAND value for $\mn$. This number of events corresponds to 3,000 days of Daya Bay data, see Table \ref{t:events} .   This figure demonstrates that the Daya Bay experiment can put  an upper limits on $\Delta m^2_{21}$ of approximately 2 times the KamLAND central value or smaller, assuming the systematic uncertainties  are smaller than the statistical uncertainties shown here.  
The dotted line is the two term approximation to the disappearance probability, see eq. (\ref{eqn:taylor}). }
\label{fig:prob}
\end{figure}

In Fig. \ref{fig:prob}, we show the electron anti-neutrino disappearance probability as function of $L/E$, keeping the quantity given in eq. (\ref{eq:fix2}) fixed, as we vary $\Delta m^2_{21}$ in multiples of $7.5 \times 10^{-5}$ eV$^2$.  Note that if $\Delta m^2_{21} > 3 \times 10^{-4}$ eV$^2$ then there is no minimum\footnote{For $\Delta_{21} < 1$, so that  $\sin \Delta_{21} \approx \Delta_{21}$, one can find the minima by finding $\Delta_{ee}$ such that,
\begin{equation} 
\frac{\mn}{\Delta m^2_{ee}} = -\frac{2\tan \theta_{13}}{ \sin 2 \theta_{12} }  \frac{ \sin 2 \Delta_{ee} }{2 \Delta_{ee} }. \nonumber
\end{equation}  
This eq. has no solutions if $\mn > 0.15 \, \mt$ or $\sim$\,4 times the KamLAND central value.}  around $L/E \approx 0.5$ km/MeV.  The red points with error bars, represents the statistical uncertainties for a detector 1.6~km from a single reactor core which has $9 \times 10^5$ events. Clearly, an experimental setup with this number of events in the far detector, 1.6~km from a reactor core, will be able to set an upper limit smaller than 3 times the KamLAND central value for  $\Delta m^2_{21}$ assuming systematic uncertainties are no larger than the statistical uncertainties.  A lower limit on $\Delta m^2_{21}$ will be challenging.

In the rest of this paper, we report on a simulation of the setups for Daya Bay and RENO experiments, to estimate the constraints these experiments can place on $\Delta m^2_{21}$.

\section{\label{sec:simulation} Simulations for Daya Bay and RENO using GLoBES}

\begin{table*}
\begin{center}
\caption{\label{t:events} 
$L_{\rm eff}$ and observed IBD $\overline{\nu}_{e}$ rates for Daya Bay and RENO derived from the GLoBES settings used in this study.\\
}
\begin{tabular*}{0.8\textwidth}{@{\extracolsep{\fill}} c c c c}
\hline
\hline
                   &         & Daya Bay & RENO \\
\hline
$L_{\rm eff}$ (m) & Near   &  (400.4, 512.6)  & 367.0 \\
                   & Far     &  1610  &  1440 \\
\hline
IBD $\overline{\nu}_{e}$ rate & Near  &  (1320, 1195)  &  617.2  \\
 (/day)         & Far    &   297.8  &  61.35   \\
\hline
\hline
\end{tabular*}
\end{center}
\end{table*}

\begin{figure*}
\begin{center}
\includegraphics[width=0.45\textwidth]{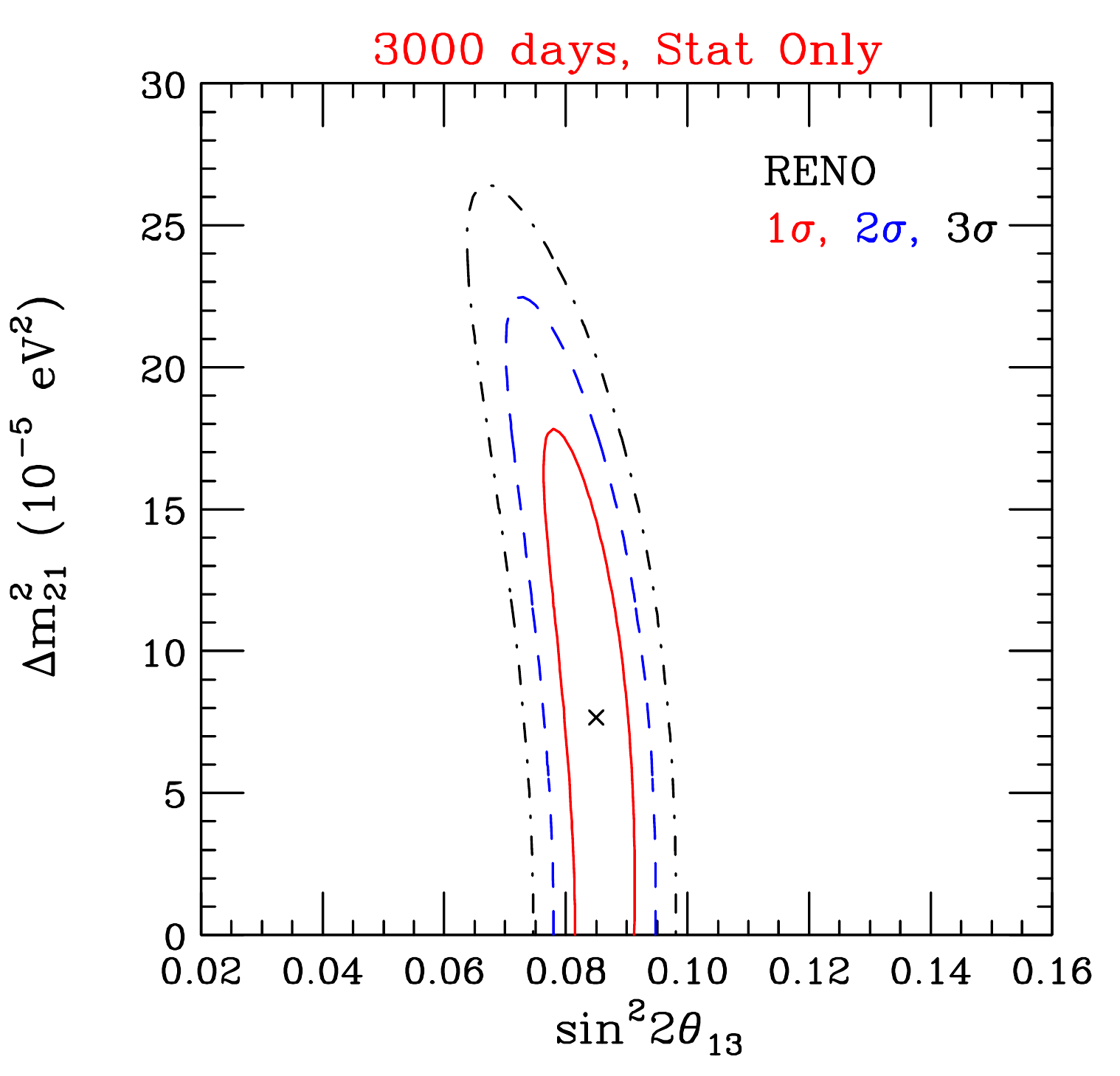}
\includegraphics[width=0.45\textwidth]{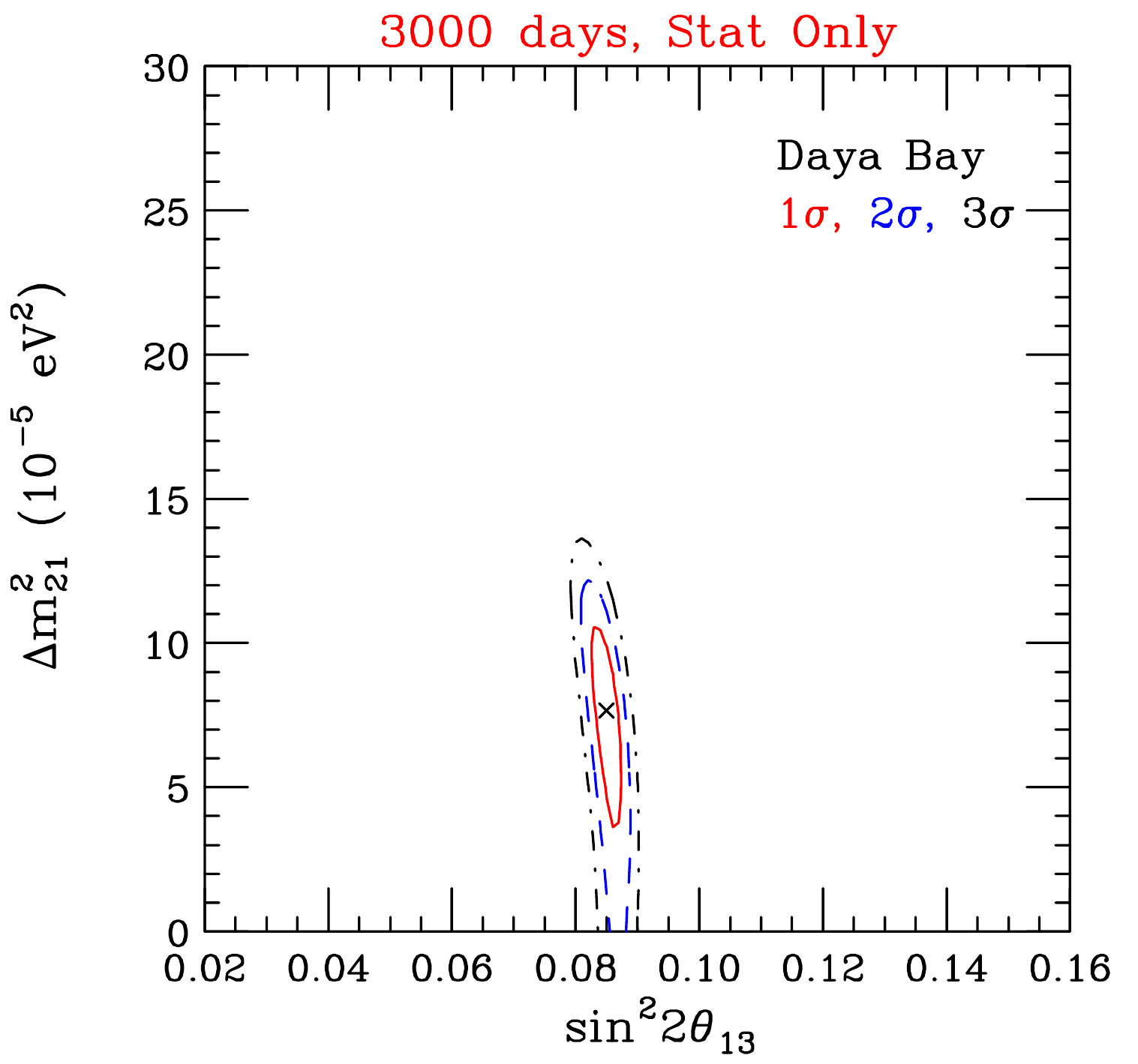}\\[3mm]
\includegraphics[width=0.45\textwidth]{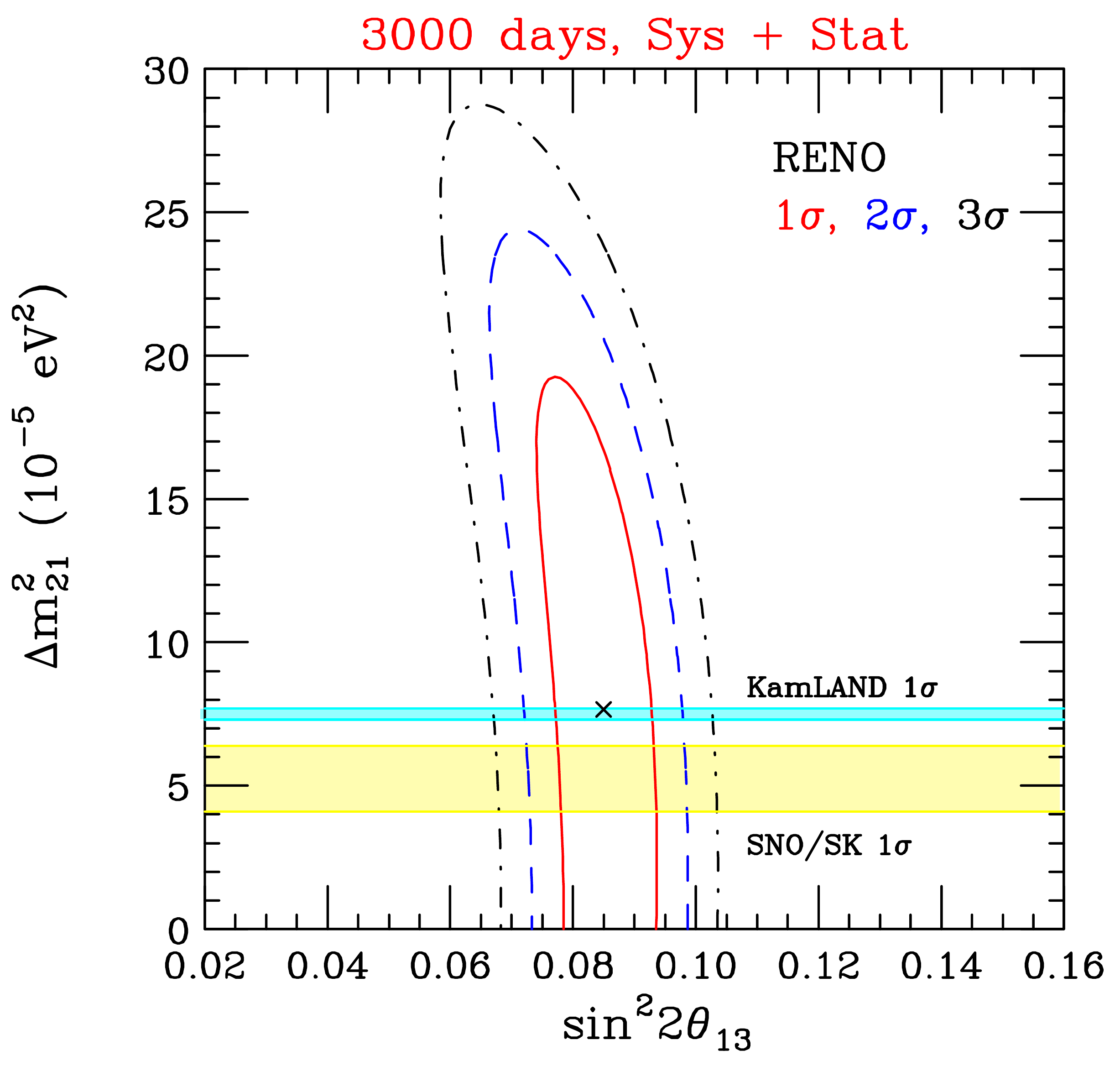}
\includegraphics[width=0.45\textwidth]{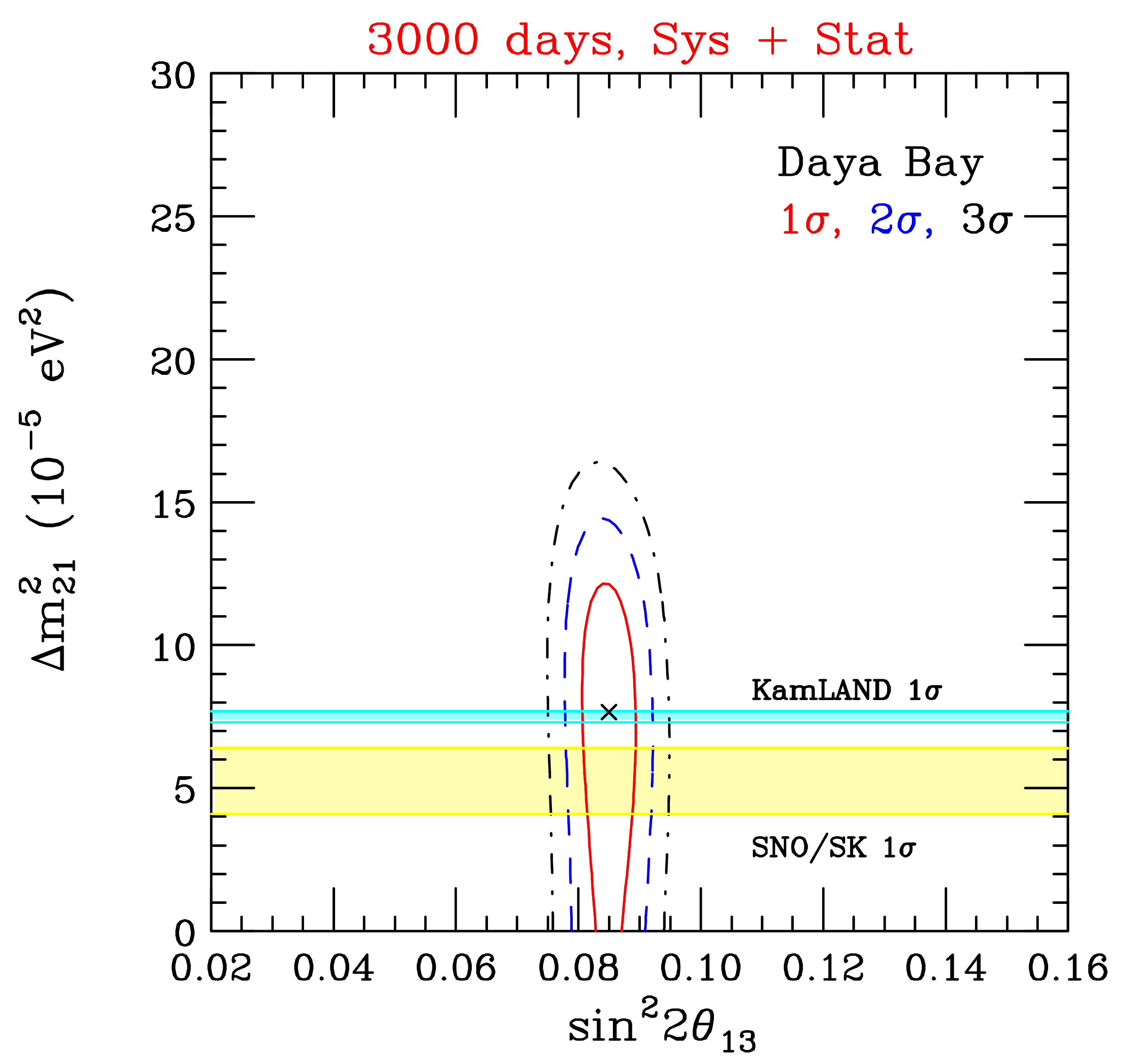}
\end{center}
\vspace{0cm}
\caption{(Color online) 
Contour plot of $\Delta m^2_{21}$ vs. $\sin^2 2 \theta_{13}$ for the RENO experiment (left column) and Daya Bay (right column)
without (top row) and with (bottom row) systematic uncertainties. 3000 live days of data with 61 \& 298 IBD $\overline{\nu}_{e}$ events/day in the far detector  were used, for RENO \& Daya Bay respectively.  
Solid (red), dashed (blue) and dotdashed (black) lines represent 1$\sigma$, 2$\sigma$, and 3$\sigma$ (2 dof) allowed regions, respectively. 
The point  ``$\times$''  is the input for the simulation given by eq. \ref{eq:fitpt}. 
In the bottom row, we also show the 1$\sigma$ uncertainty band on $\mn$ from KamLAND (cyan) and SNO/SK (yellow), see eq. \ref{eq:kamland} and \ref{eq:sk+sno}.  The value of $\sin^2 \theta_{12}$ is held fixed for this analysis, for a discussion on varying $\theta_{12}$ see Appendix A.
}
\label{f:3000d}
\end{figure*}

Our sensitivity study on $\Delta m^2_{21}$ for the short baseline reactor experiments, Daya Bay and RENO, is performed
using GLoBES~\cite{globes}.
In this study 3000 live days of data are assumed for both experiments
and systematic uncertainties are taken into account as 
described in ~\cite{db_1230d} for Daya Bay and ~\cite{reno_500d} for RENO. 
Table~\ref{t:events} lists the effective baselines, $L_{\rm eff}$, and the number of observed IBD $\overline{\nu}_{e}$ events per day used.

To find the best fit values of $\Delta m^2_{21}$ and sin$^{2}(2\theta_{13})$, a $\chi^2$ formalism with pull parameters
is constructed using the far-to-near ratio method to cancel out correlated systematic uncertainties. The $\chi^2$ is given by
\vspace{2mm}
\begin{eqnarray}
 \chi^2  & = & \sum_{i=1}^{\rm{N}_{\rm{bins}}} \frac{ (O_i^{\rm{F/N}} - X_i^{\rm{F/N}})^2 }{U_i^{\rm{F/N}}}
+ \sum_{r=1}^{6} \left( \frac{f^r}{\sigma_{\rm{flux}}^r} \right)^2 
                          + \left( \frac{\epsilon}{\sigma_{\rm{eff}}} \right)^2
                                 \nonumber \\ & &
  \quad    \quad \quad                     + \left( \frac{s}{\sigma_{\rm{scale}}} \right)^2 
+ \sum_{d={\rm N, F}} \left( \frac{b^{d}}{\sigma_{\rm{bkg}}^{d}} \right)^2,
\end{eqnarray}
where,
\begin{itemize}
\item $O_i^{\rm{F/N}}$ is the observed far-to-near ratio of IBD $\overline{\nu}_{e}$ events in the $i$-th $E_{\overline{\nu}}$ bin,
\item $X_i^{\rm{F/N}} = X_i^{\rm{F/N}} (f^r, \epsilon, s, b^d; \theta_{13}, \Delta m_{21}^2) $ is the expected far-to-near ratio of IBD $\overline{\nu}_{e}$ events
for a given $\Delta m_{21}^2$ and $\theta_{13}$ pair,
\item $U_i^{\rm{F/N}}$ is the statistical uncertainty of $O_i^{\rm{F/N}}$,
\item $f^r, \epsilon, s,$ and  $b^d$ are pull parameters for systematic uncertainties of 
neutrino flux ($\sigma_{\rm{flux}}^r$), detection efficiency ($\sigma_{\rm{eff}}$), 
energy scale ($\sigma_{\rm{scale}}$), and background ($\sigma_{\rm{bkg}}^{d}$), respectively. 
\end{itemize}
The indices $r$ and $d$ represent $r$-th reactor and $d$-th detector, respectively. Both Daya Bay and RENO have six reactors.
For Daya Bay, two near detector sets (N$_{1}$ and N$_{2}$) are used in the last pull term of the $\chi^2$ 
due to their differences in the baselines, backgrounds, and systematic uncertainties~\cite{db_1230d}.
As a cross check of our simulations we have reasonably well reproduced the $\Delta m^2_{ee}$ vs. $\sin^2 2 \theta_{13}$ sensitivity curves for both experiments.

True values used in the signal simulation are 
\begin{eqnarray}
\sin^{2}\theta_{12} = 0.304, &\quad & \Delta m^2_{21} = 7.65\times 10^{-5}~{\rm eV}^2, \nonumber \\
\sin^{2}(2\theta_{13}) = 0.085, &&\Delta m^2_{31} = 2.50\times 10^{-3}~{\rm eV}^2.
\label{eq:fitpt}
\end{eqnarray}
For this analysis the value of $\sin^2 \theta_{12}$ is held fixed. For discussion on varying $\theta_{12}$, see Appendix A.

To minimize the $\chi^2$, expected values for different pairs of $\Delta m^2_{21}$ and sin$^{2}(2\theta_{13})$
are compared to the simulated signal $\overline{\nu}_{e}$ data from 1.8 to 8~MeV with 31 energy bins.

Figure~\ref{f:3000d} shows the results of our simulation for contour plots of $\Delta m^2_{21}$ vs. sin$^{2}(2\theta_{13})$ 
sensitivities using 3000 live days of data for RENO and Daya Bay, respectively, without (top) and with (bottom) systematic uncertainties. 
Adding systematic uncertainties effects RENO less than Daya Bay, because after 3,000 days of data taking,  Daya Bay has $\approx$ 5 times more events in the far detector(s) than RENO, see Table \ref{t:events}.
Clearly, both of these experiments\footnote{In \cite{An:2016luf}, Fig. 3, Daya Bay gives constraints on a 3+1 sterile neutrino model with 600 days of data.  These constraints can be re-interpreted as a constraint on $\Delta m^2_{21}$  and the result is slightly better than 3 times the KamLAND value. Validating our conclusion of 2 - 3 times the KamLAND value is achieveable.} can constrain $\mn$ to be less than two to three times the KamLAND central  value, i.e. $\mn < 15-22 \times 10^{-5}$ eV$^2$.
Setting a lower limit on $\mn$ maybe possible with Daya Bay if improvements in their systematic uncertainties, over those used for this simulation, can be achieved.
We encourage both Daya Bay and RENO to perform a measurement of $\mn$ using their more precise information on their experiments.

\section{\label{sec:summary} Conclusion}

We have argued that Daya Bay and RENO can add to the information of the solar mass squared difference, $\mn$, now.
A simulation study for these experiments was performed with and without systematic uncertainties using GLoBES. 
We have found that $\Delta m^2_{21}$ can be reasonably well constrained by Daya Bay 3000 live days of data to be less than twice the KamLAND central value at the 95\% CL.
Without systematic uncertainties Daya Bay can exclude $\Delta m^2_{21} = 0$ with 1$\sigma$  confidence level
but when current systematic uncertainties are included only an upper bound can be set. 
Until JUNO measures $\Delta m^2_{21}$ with great precision in the middle of next decade,
we expect the $\Delta m^2_{21}$ measurement by Daya Bay can play an important role for the leptonic CP violation measurement by T2K and NOvA and  provides an  important consistency check on the 3 flavor massive neutrino paradigm.
In this study we fixed $\theta_{12}$ value since the uncertainty from $\theta_{12}$ variation is relatively small compared to the systematic uncertainties as discussed in Appendix A. A truly realistic simulation and a true measurement of $\Delta m^2_{21}$ can only be performed by the short baseline reactor experiments, Daya Bay and RENO.

\begin{acknowledgments}

This work (SHS) was supported by the National Research Foundation of Korea (NRF) grant funded by the Korea government (MSIT)
(No. 2017R1A2B4012757 and IBS-R016-D1-2018-b01).

This manuscript has been authored (SJP) by Fermi Research Alliance, LLC under Contract No.~DE-AC02-07CH11359 with the U.S. Department of Energy, Office of Science, Office of High Energy Physics.

This project (SJP) has received funding/support from the European Union's Horizon 2020 research and innovation programme under the Marie Sklodowska-Curie grant agreement No 690575 \& No 674896.
\end{acknowledgments}

\appendix
\section{$\Delta m^2_{21}$ sensitivity to variation of $\theta_{12}$}
For the Daya Bay and RENO experiments, the disappearance probability is well approximated by
\begin{eqnarray}
& & 1- P(\bar{\nu}_e \rightarrow \bar{\nu}_e)  \approx  \nonumber \\
& &    \sin^2 2 \theta_{13}  \sin^2 \Delta_{ee} + 
 \cos^4\theta_{13} ( \sin 2 \theta_{12} \Delta_{21})^2,
 \label{eqn:taylor2}
\end{eqnarray}
and therefore these experiments are only sensitive to variables $\sin^2 \theta_{13}$, $\Delta m^2_{ee}$ and the product $(\sin 2 \theta_{12} \Delta m^2_{21})$.
That is, there is a degeneracy between $\sin 2 \theta_{12}$ and $ \Delta m^2_{21}$ as long as the product is same.  So in principle the vertical axes of Fig. \ref{f:3000d} could be replaced by
\begin{eqnarray}
 \Delta m^2_{21}  \quad   \Longrightarrow  \quad   \Delta m^2_{21} ( \sin 2 \theta_{12} / 0.92)
 \end{eqnarray}
where 0.92 is the value of  $\sin 2 \theta_{12}$ used to produce these figures.

By applying the 3 $\sigma$ level allowed range of $\sin^2 \theta_{12}$ for KamLAND, i.e. [0.20, 0.42], wider than for the SK/SNO, i.e. [0.27, 0.36], see \cite{Gando:2010aa, Aharmim:2011vm, Abe:2010hy, Esteban:2016qun}, the $\Delta m^2_{21}$ measurement would have been affected by $\sim$ 13\% or less, as the dependence comes from $\sin 2 \theta_{12}$ not $\sin^2 \theta_{12}$,
Therefore the systematic uncertainties of the experiments on $\Delta m^2_{21}$ measurement, is much larger than the uncertainty from the variation of $\theta_{12}$.


In conclusion, for the short baseline reactor experiments, variation of $\theta_{12}$ has relatively small impact on the measurement of 
$\Delta m^2_{21}$ as well as $\sin^2 2 \theta_{13}$, $\Delta m^2_{ee}$
\footnote{If the value of $\Delta m^2_{31}$ and/or  $\Delta m^2_{32}$  is extract from the measurement of $\Delta m^2_{ee}$ then the sensitivity to $\sin ^2 \theta_{12}$ is increased as well as sensitivity to the mass ordering.}.

\end{document}